\documentclass[12pt]{article}
\usepackage{graphicx}
\usepackage{url}

\newcommand\pubdate{February 24, 2014}

\def\support{{}}

\def\Title#1{\begin{center} {\Large #1 } \end{center}}
\def\Author#1{\begin{center}{ \sc #1} \end{center}}

\newcommand\pubblock{\rightline{\begin{tabular}{l} \\
         \pubdate  \end{tabular}}}
\newenvironment{Abstract}{\begin{quotation}  }{\end{quotation}}
\newenvironment{Presented}{\begin{quotation} \begin{center} 
             PRESENTED AT\end{center}\bigskip 
      \begin{center}\begin{large}}{\end{large}\end{center} \end{quotation}}
\def\Acknowledgements{\bigskip  \bigskip \begin{center} \begin{large}
             \bf ACKNOWLEDGEMENTS \end{large}\end{center}}

\def\beq{\begin{equation}}
\def\eeq#1{\label{#1}\end{equation}}
\def\eeqn{\end{equation}}

\def\beqa{\begin{eqnarray}}
\def\eeqa#1{\label{#1}\end{eqnarray}}
\def\eeqan{\end{eqnarray}}

\let\bar=\overbar

\def\etal{{\it et al.}}

\def\Dslash{\not{\hbox{\kern-4pt $D$}}}
\def\dslash{\not{\hbox{\kern-2pt $\del$}}}

\def\msb{{\bar{\ssstyle M \kern -1pt S}}}

\newcommand{\keVnr}{keV$\mathrm{_{nr}}$} 
\newcommand{\keVee}{keV$\mathrm{_{ee}}$} 

\newcommand{\mus}{\ensuremath{\mu{}s}} 
\newcommand{\mwe}{mwe} 

\newcommand{\scix}[3][]  
{ 
 \ifthenelse{\equal{#2}{}} 
  {} 
  { 
   {#2} 
  } 
 \ifthenelse{\equal{#2}{}\OR\equal{#3}{}} 
  {} 
  { 
   {\hspace{-0em} \times \hspace{-0em}} 
  } 
 \ifthenelse{\equal{#3}{}} 
  {} 
  {10^{#3}}   
 \ifthenelse{\equal{#1}{}}
 {}
 {\,\mathrm{#1}} 
}

\begin{document}
\begin{titlepage}
\pagenumbering{gobble}
\pubblock

\makeatletter
\renewcommand*{\@fnsymbol}[1]{\ensuremath{\ifcase#1\or 1\or 2\or 3\or 4\or 5\or 6\or 7\or 8\or 9\or 10\or 11\or 12\or 13\or 14\or 15\or 16\or 17\or 18\or 19\or 20\or *\or \dagger\or , \else\@ctrerr\fi}}
\makeatother

\renewcommand*{\thefootnote}{\fnsymbol{footnote}}

\vfill
\Title{A Detailed Look at the First Results from the Large Underground Xenon (LUX) Dark Matter Experiment}
\vfill
\Author{ Matthew Szydagis\footnotemark[1]\footnote[21]{Corresponding Author: mmszydagis@ucdavis.edu}, \\on behalf of the LUX collaboration \\(D.S.~Akerib\footnotemark[2], H.M.~Ara\'{u}jo\footnotemark[3], X.~Bai\footnotemark[4], A.J.~Bailey\footnotemark[3], J.~Balajthy\footnotemark[5], E.~Bernard\footnotemark[6], A.~Bernstein\footnotemark[7], A.~Bradley\footnotemark[2], D.~Byram\footnotemark[8], S.B.~Cahn\footnotemark[6], M.C.~Carmona-Benitez\footnotemark[9], C.~Chan\footnotemark[10], J.J.~Chapman\footnotemark[10], A.A.~Chiller\footnotemark[8], C.~Chiller\footnotemark[8], T.~Coffey\footnotemark[2], A.~Currie\footnotemark[3], L.~de\,Viveiros\footnotemark[11], A.~Dobi\footnotemark[5], J.~Dobson\footnotemark[12], E.~Druszkiewicz\footnotemark[13], B.~Edwards\footnotemark[6], C.H.~Faham\footnotemark[14], S.~Fiorucci\footnotemark[10], C.~Flores\footnotemark[1], R.J.~Gaitskell\footnotemark[10], V.M.~Gehman\footnotemark[14], C.~Ghag\footnotemark[15], K.R.~Gibson\footnotemark[2], M.G.D.~Gilchriese\footnotemark[14], C.~Hall\footnotemark[5], S.A.~Hertel\footnotemark[6], M.~Horn\footnotemark[6], D.Q.~Huang\footnotemark[10], M.~Ihm\footnotemark[16], R.G.~Jacobsen\footnotemark[16], K.~Kazkaz\footnotemark[7], R.~Knoche\footnotemark[5], N.A.~Larsen\footnotemark[6], C.~Lee\footnotemark[2], A.~Lindote\footnotemark[11], M.I.~Lopes\footnotemark[11], D.C.~Malling\footnotemark[10], R.~Mannino\footnotemark[17], D.N.~McKinsey\footnotemark[6], D.-M.~Mei\footnotemark[8], J.~Mock\footnotemark[1], M.~Moongweluwan\footnotemark[13], J.~Morad\footnotemark[1], A.St.J.~Murphy\footnotemark[12], C.~Nehrkorn\footnotemark[9], H.~Nelson\footnotemark[9], F.~Neves\footnotemark[11], R.A.~Ott\footnotemark[1], M.~Pangilinan\footnotemark[10], P.D.~Parker\footnotemark[6], E.K.~Pease\footnotemark[6], K.~Pech\footnotemark[2], P.~Phelps\footnotemark[2], L.~Reichhart\footnotemark[15], T.~Shutt\footnotemark[2], C.~Silva\footnotemark[11], V.N.~Solovov\footnotemark[11], P.~Sorensen\footnotemark[7], K.~O'Sullivan\footnotemark[6], T. Sumner\footnotemark[3], D.~Taylor\footnotemark[18], B.~Tennyson\footnotemark[6], D.R.~Tiedt\footnotemark[4], M.~Tripathi\footnotemark[1], S.~Uvarov\footnotemark[1], J.R.~Verbus\footnotemark[10], N.~Walsh\footnotemark[1], R.~Webb\footnotemark[17], J.T.~White\footnotemark[17]\footnote[22]{deceased}, M.S.~Witherell\footnotemark[9], F.L.H.~Wolfs\footnotemark[13], M.~Woods\footnotemark[1], C.~Zhang\footnotemark[8])\support}
\vfill

\begin{tiny}

\noindent \footnotemark[1]\it{University of California Davis, Dept. of Physics, One Shields Ave., Davis CA 95616, USA}\\
\footnotemark[2]Case Western Reserve University, Dept. of Physics, 10900 Euclid Ave, Cleveland OH 44106, USA\\
\footnotemark[3]Imperial College London, High Energy Physics, Blackett Laboratory, London SW7 2BZ, UK\\
\footnotemark[4]South Dakota School of Mines and Technology, 501 East St Joseph St., Rapid City SD 57701, USA\\
\footnotemark[5]University of Maryland, Dept. of Physics, College Park MD 20742, USA\\
\footnotemark[6]Yale University, Dept. of Physics, 217 Prospect St., New Haven CT 06511, USA\\
\footnotemark[7]Lawrence Livermore National Laboratory, 7000 East Ave., Livermore CA 94551, USA\\
\footnotemark[8]University of South Dakota, Dept. of Physics, 414E Clark St., Vermillion SD 57069, USA\\
\footnotemark[9]University of California Santa Barbara, Dept. of Physics, Santa Barbara, CA, USA\\
\footnotemark[10]Brown University, Dept. of Physics, 182 Hope St., Providence RI 02912, USA\\
\footnotemark[11]LIP-Coimbra, Department of Physics, University of Coimbra, Rua Larga, 3004-516 Coimbra, Portugal\\
\footnotemark[12]SUPA, School of Physics and Astronomy, University of Edinburgh, Edinburgh, EH9 3JZ, UK\\
\footnotemark[13]University of Rochester, Dept. of Physics and Astronomy, Rochester NY 14627, USA\\
\footnotemark[14]Lawrence Berkeley National Laboratory, 1 Cyclotron Rd., Berkeley, CA 94720, USA\\
\footnotemark[15]Department of Physics and Astronomy, University College London, Gower Street, London WC1E 6BT, UK\\
\footnotemark[16]University of California Berkeley, Department of Physics, Berkeley, CA 94720, USA\\
\footnotemark[17]Texas A \& M University, Dept. of Physics, College Station TX 77843, USA\\
\footnotemark[18]South Dakota Science and Technology Authority, SURF, Lead, SD 57754, USA\\

\end{tiny}

\begin{Abstract}
LUX, the world's largest dual-phase xenon time-projection chamber, with a fiducial target mass of 118~kg and 10,091 kg-days of exposure thus far, is currently the most sensitive direct dark matter search experiment. The initial null-result limit on the spin-independent WIMP-nucleon scattering cross-section was released in October~2013, with a primary scintillation threshold of 2~phe, roughly 3~\keVnr~for LUX. The detector has been deployed at the Sanford Underground Research Facility (SURF) in Lead, South Dakota, and is the first experiment to achieve a limit on the WIMP cross-section lower than $10^{-45}$~cm$^{2}$. Here we present a more in-depth discussion of the novel energy scale employed to better understand the nuclear recoil light and charge yields, and of the calibration sources, including the new internal tritium source. We found the LUX data to be in conflict with low-mass WIMP signal interpretations of other results.
\end{Abstract}
\vfill
\begin{Presented}
Symposium on Cosmology and Particle Astrophysics\\
Honolulu, Hawai'i, November 12--15, 2013
\end{Presented}
\vfill
\end{titlepage}
\def\thefootnote{\fnsymbol{footnote}}
\setcounter{footnote}{0}

\pagenumbering{arabic}
\section{Introduction}

The body of indirect evidence for dark matter is extensive. It includes the best-fit model for explaining the angular power spectrum of the Cosmic Microwave Background temperature anisotropy, gravitational lensing studies, large-scale structure observations and simulations, and galactic rotation curves~\cite{Blumenthal:1984bp,Davis:1985,Clowe:2006,Hinshaw:2012aka,Ade:2013zuv}. All these point to a significant non-baryonic, cold (heavy and non-relativistic) component of matter in the Universe: $\sim$85\% of the matter, or, $\sim$27\% of the total mass-energy content of the Universe. The WIMP (Weakly Interacting Massive Particle) is a favored candidate, with many theories (for example, supersymmetry and Kaluza-Klein) providing natural candidate particles. Most direct dark matter searches are therefore geared towards finding WIMPs, which are expected to produce low-energy nuclear recoils (NR) in a detector, while electron recoils (ER) constitute a primary background to be reduced and identified~\cite{PhysRevD.31.3059, Feng:2010}.

The noble element xenon has many favorable properties as a target for direct WIMP detection experiments~\cite{araujoReview}. For such an element, deposited energy in the material is expressed in three possible channels: excitation, ionization, and heat. The most prominent channel for NR is heat, reducing the amount of energy in the first two significantly, in contrast to ER. Ionization electrons can recombine, or escape. Excitation and recombination together lead to the so-called primary scintillation signal (S1), with S1 from each source indistinguishable. Escaping ionization electrons lead to the secondary scintillation (S2) in the gas region of a two-phase detector. Lastly, noble elements are transparent to their own scintillation light because it originates in decaying molecules (excited dimers, i.e. excimers) rather than in atoms.

\section{The LUX Detector}

The LUX (Large Underground Xenon) collaboration is comprised of 18 institutions from the US, UK, and Portugal. The detector, a full treatment of which is found in~\cite{LUXNimPaper}, is a two-phase xenon time-projection chamber (TPC) with two photo-multiplier tube (PMT) arrays, top and bottom. Each has 61 PMTs with quantum efficiency (QE), i.e., probability for conversion of a photon into a photo-electron (phe), of $\sim$30--40\% (dependent on individual PMT, angle and position of incidence, and temperature) at the ultraviolet xenon scintillation wavelength ($\sim$175~nm)~\cite{LUXPMTPaper}. The time in between the S1 and S2 signals (0--324~\mus~in LUX) provides the depth of an event, while the S2 hit pattern of light in the top PMT array provides the radial position. The ratio of S2 to S1, versus S1, is the basis of the NR versus ER discrimination of backgrounds. Because liquid xenon is dense ($\sim$3~g/cm$^{3}$) and a high-Z material, it is good at self-shielding of gamma and neutron backgrounds. This makes fiducialization and multiple-scattering rejection useful and powerful techniques in LUX.

To minimize the background cosmic ray flux, the detector is deployed at a depth of 4300~\mwe~(4850~ft.) at SURF (Sanford Underground Research Facility), the former location of the Homestake gold mine, where the ``solar neutrino problem" was first discovered~\cite{lab1,lab2,lab3}. The TPC is housed within a low-background titanium cryostat~\cite{TiPaper}, inside of a water tank 6.1~m (height) by 7.6~m (diameter), which reduces the background. LUX contains 370~kg gross, but  250~kg in the active region, defined as the electron drift region (cathode to gate grid)~\cite{LUXNimPaper}. An inner 118~kg is defined as the fiducial mass for analysis. The distance from the cathode close to the bottom of the TPC to the gate near the top is 48~cm, and 47~cm is the approximate diameter of the dodecagonal active region, within which a 181~V/cm electric field is set up for the purpose of drifting the ionization electrons upwards to the S2 region at top. The finite electron absorption length, caused by residual electronegative impurities, was 87--134~cm over the course of the run. Given the 1.51~mm/\mus~electron drift speed, this corresponds to a range in electron ``lifetime" of $\sim$500--900~\mus. The electron extraction field was 6.0~kV/cm in the gas (3.1 in the liquid), resulting in 65$\pm$4\% extraction efficiency. An S2 analysis threshold of 200~phe was implemented, driven by a conservative treatment of the position resolution of wall events, and the radial fiducial cut, ensuring no edge events would be reconstructed within the fiducial volume. Given the size of a single electron signal (25$\pm$7~phe) this corresponds to a mean of 8 extracted electrons approximately.

\section{Light Collection and Light Yield}

The interior walls of the detector as well as the spaces between the PMTs are covered with PTFE. Measurements of PTFE reflectivity in liquid xenon in the UV and best-fit Geant4~\cite{G4} optical (photon propagation) Monte Carlo simulations of LUX agree that it is $>$95\%~\cite{LUXRun02,Sz_IDM}. We determined that the detection efficiency for an S1 photon originating at the center of the detector is 14$\pm$1\%, a number which combines both the geometric light collection efficiency resulting from a finite photon absorption length and poorly reflective stainless steel grid wires, plus the QEs of the PMTs. Photon detection efficiency varies from 11--17\% between the top and bottom of the TPC respectively. Total internal reflection at the liquid/gas interface causes most S1 light to be detected by the bottom PMT array. The position dependence of the S1 was ``corrected" by flat-fielding the detector response, using the center as the reference point. The photon detection efficiency was cross-checked using multiple calibration sources, as well as different methods, which were all in agreement: division of observed yield by the expected yield from the Noble Element Simulation Technique model (NEST, described in greater detail later in this section)~\cite{Szydagis2013}, optimization of the energy resolution for high-energy mono-energetic ER calibration peaks~\cite{Conti,ZIII}, as well as the prediction based on the reflectivities and photon absorption lengths. By comparing to NEST, we estimate the zero-field yield to be 8.8~phe/keV for a 122~keV gamma. This value can be compared to XENON100, where the stated light yield of 2.28~phe/keV at 530~V/cm corresponds to approximately 3.9~phe/keV at null field using their assumed multiplicative factor of 0.58 for the size of the S1 signal with respect to 0~V/cm~\cite{Xe100}.

The traditional energy estimator in two-phase xenon experiments is based on S1 and defined with reference to the 122 keV $^{57}$Co line~\cite{Xe100,Xe1002011}, but this gamma ray does not penetrate large detectors effectively.~Furthermore, given the well-understood and energy-dependent partition of energy between the light and charge channels, an estimator based on S1 and S2 improves resolution by mitigating recombination and light collection fluctuations~\cite{Conti}. We opt for such a combined estimator, defined with reference to initial quanta, where for a nuclear recoil the reconstructed energy in units of~\keVnr~is defined as~\cite{DahlThesis}

\begin{equation}
E = \mathcal{L}^{-1} (N_{\gamma} + N_{e-}) W, \quad W = 13.7~\pm~0.2~\mathrm{eV}.
\end{equation}

Energy is proportional to the average work function for production of either a photon or an electron $W$, and is a linear combination of $N_{\gamma}$ and $N_{e-}$, the numbers of primary photons and electrons generated respectively, calculated as follows~\cite{SorDahl2011}:
\begin{equation}
N_{\gamma} = \mathrm{S1_{c}} / g_{1}, \quad N_{e-} = \mathrm{S2_{c}} / (\epsilon  g_{2}),
\end{equation}
\begin{equation}
g_{1}=0.14~\mathrm{phe/photon}, \quad \epsilon=0.65, \quad g_{2}=24.55~\mathrm{phe/electron},
\end{equation}
where S1$_{c}$ is the S1 signal in phe, corrected to the detector center, S2$_{c}$ is the S2 signal in phe corrected instead to the liquid surface, $\epsilon$ is the electron extraction efficiency, $g_{1}$ is the S1 photon detection efficiency, and $g_{2}$ is the average single electron signal size. The position corrections in three dimensions for both the S1 and S2 are accomplished utilizing over one million $^{83\mathrm{m}}$Kr~\cite{Kastens} events distributed uniformly throughout the detector volume, and recorded periodically over the entire course of the run. In order of increasing energy, the energy scale was calibrated using tritium, $^{83\mathrm{m}}$Kr, and cosmogenically activated xenon lines at 164 and 236~keV. The factor $\mathcal{L}$ is the ratio of energy channeled into ionization plus excitation to the total amount of energy deposited, including heat. The assumption that $\mathcal{L} = 1$ for ER forms the basis for a good match to past data~\cite{Szydagis2011}, while for NR an energy-dependent $\mathcal{L}$ describes the reduction in light and charge yields with respect to ER due to the greater propensity for it to generate further NR. The Hitachi-corrected Lindhard factor is assumed in LUX ($k$ parameter of 0.110 instead of 0.166 as in standard Lindhard theory for xenon). It provides a good yet low-yield and conservative match to the LUX NR yields as well as NR data from past experiments~\cite{SorDahl2011,Szydagis2011}.

The absolute scintillation and ionization yields for NR as a function of energy and field are modeled using NEST~\cite{Szydagis2013}, which is based on the canon of existing experimental data~\cite{Szydagis2011}, including the Ph.D. thesis data of Dahl, from five different fields (60, 522, 876, 1951, 4060 V/cm)~\cite{DahlThesis}. From these data it was possible to extract the energy-dependent light suppression factors for electric field. With a higher field, recombination probability is decreasing. Thus, light yield decreases at the expense of increased charge. We estimate that NR light production at the LUX field is $\sim$0.8 times its zero-field value (as defined by the Plante~\etal~data~\cite{Plante2011}). NEST is not only low in S1 yield and thus conservative, but also predictive, matching the LUX data $a~priori$, treating only detector parameters as free, such as photon detection efficiency, and not absolute yields. The charge yield assumed using NEST is also lower than that assumed by any other experiment, even accounting for the lower LUX field~\cite{Sorensen2010}.

\section{Pulse Finding Efficiencies}

\begin{figure}[h]
\begin{center}
\includegraphics[width=0.95\textwidth]{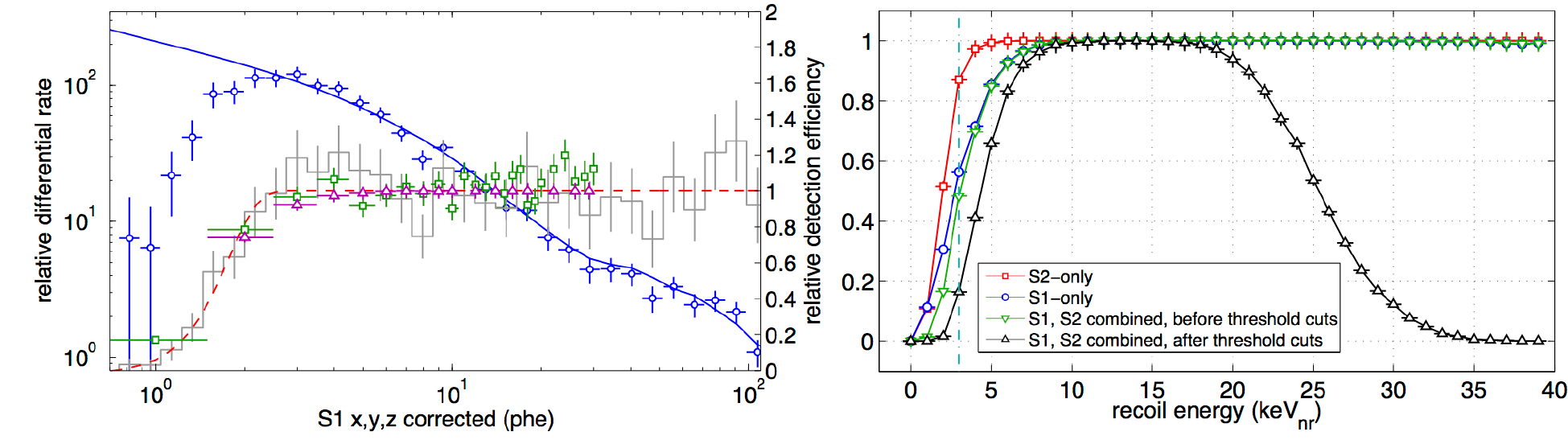}
\caption{\label{Efficiencies} {\it Left}: Comparison of LUX AmBe data (blue circles) with NEST simulation (blue line), showing excellent agreement above the 2~phe threshold (left axis).  The gray histogram and fitted dashed red line show the relative efficiency for detection of nuclear recoils from AmBe data (right axis), relative in the sense that overall normalization must be determined separately, as explained in the text. Overlaid are the ER detection efficiency from tritium data (green squares), applied to the ER background model in the profile likelihood analysis, and the efficiency from full detector NR simulations treated as real data, in terms of the digitized MC-truth S1 phe (purple triangles), applied to the WIMP signal model. The efficiency calculation here does not include S1 or S2 area thresholds. {\it Right}: WIMP detection efficiency as a function of nuclear recoil energy for events with a corrected S1 between 2 and 30~phe and an S2 signal greater than 200~phe (black triangles). This efficiency is used directly in the profile likelihood analysis. In addition, we show the efficiency for individually detecting an S2 (red squares) or S1 (blue circles) signal, respectively, without the application of any analysis thresholds. The detection efficiency for single scatter events (again applying no threshold cuts), shown by the green triangles, clearly demonstrates the dominant impact of the S1-only efficiency.  The cyan dashed line indicates the threshold in~\keVnr~below which we assume neither light nor charge response in the context of the WIMP dark matter signal model.}
\end{center}
\end{figure}

Pulse finding and classification efficiency was studied in great detail using the AmBe and $^{252}$Cf calibrations (effective at producing low-energy NR), tritiated ($^{3}$H) methane (a beta emitter used as the primary low-energy ER calibration), and full Monte Carlo simulations formatted and processed in the same fashion as real data~\cite{LUXSim}. First, the NEST AmBe light yield was assumed for a simplified 100\%-efficiency simulation, in order to derive the efficiency for detection of single-scatter NR events (WIMP-like) by comparing the simulated spectrum to neutron calibration data. Excellent agreement was observed compared to the efficiency for tritium, as well as compared to efficiency observed in full simulations where it is already present because of S1 and S2 size and pulse shape modeling~\cite{NESTmock}. These comparisons were performed versus S1 area and not combined energy, lowering the 200~phe S2 threshold to 100 in order to compare NR and ER, mitigating the effect of their different yields (Figure~\ref{Efficiencies}, left). Once photons are created, the particle type which generated them is irrelevant. A hand scan of events was performed which included verifying the asymptote to 100\% relative efficiency. The energy-independent component of efficiency for detection of 1 S1 and 1 S2 (WIMPs do not multiply scatter) was found to be 98\%, and the expected number of tritium events confirmed absolute efficiency was near 100\%.

Efficiency can be broken down into S1 pulse identification and classification, S2, and the efficiency for simultaneous identification in one event, and can be plotted as a function of energy (Figure~\ref{Efficiencies}, right) or S1 size (not S2 size, because the S1 pulse finding efficiency for low-energy events dominates). The relevant analysis cuts were a 2-fold PMT coincidence, a minimum S1$_{c}$ area of 2~phe, and a minimum S2 (raw, not position-corrected) area of 200~phe. The 2-fold coincidence requirement and 2~phe area threshold are not identical, due to position correction of the S1 signal to the detector center and the typical single phe resolution of 30\% for the LUX PMTs (R8778)~\cite{LUXPMTPaper}. For events which meet the coincidence requirement, 2~phe must truly have been present (digitally). However, the events passing that requirement but failing to exceed the area threshold (analog) are conservatively cut. For the LUX S1 and S2 yields a true-energy 3~\keVnr~nuclear recoil would produce 2.0~phe S1 on average (only a mean because of finite energy resolution, and lack of 1:1 correspondence between energy and S1~\cite{Sorensen2012}). It would also produce over 200~phe S2, so the given S1 and S2 thresholds ensure that no model for ``sub-threshold fluctuations" is required to derive the final limit. Furthermore, the limit calculation conservatively assumed that NR produces no S1 or S2 below 3~\keVnr~(true energy, not reconstructed), the lowest energy point for which empirical data (from angle-tagged neutron-scattering measurements) existed demonstrating a non-zero yield~\cite{Plante2011}. Recent results from an $in~situ$ calibration performed utilizing the mono-energetic neutrons of a DD neutron generator~\cite{Verbus2014} underscore the validity of the LXe signal model assumptions at, and above, the unphysical 3~\keVnr~cutoff adopted for the initial LUX WIMP limit. This new work also provides signal calibrations below 3~\keVnr~that may be used to further extend the sensitivity of LUX.

\section{WIMP Search Data and Calibrations}

Over the course of April through August~2013, 85.3 live-days of background, WIMP-search data were taken. Calibrations were performed throughout the run. The mean background event rate, driven by ER, was measured to be 3.6$\pm$0.3~mDRU (counts per keV-kg-day) in the energy and volume regions of interest, the lowest achieved for a xenon TPC~\cite{ZIII,Xe100,Xe1002011}. An intrinsic background for xenon is krypton, as $^{85}$Kr is a beta emitter; LUX achieved a level of 3.5$\pm$1~ppt~Kr measured before the start of the run and $in~situ$~\cite{krRemoval,dobi,ChangAPS}. Over the course of data-taking, cosmogenic activity caused originally by surface operation~\cite{LUXRun02} dropped by 0.5~mDRU in the 118.3$\pm$6.5~kg fiducial mass. The fiducial volume was defined by a radial cut placed at 18~cm and a cut along the drift field axis: $z=7-47$~cm included, where $z$ is defined as the upward distance from the bottom PMT array, with the liquid/gas border at $\sim$55~cm. This first LUX result was a non-blind analysis, with events triggered on S2, and the cuts made to the data were simple. Cuts applied included: detector stability in pressure, voltage, and liquid level; a single-scatter cut; energy cuts via S1 and S2 cuts; a non-S1, non-S2 area cut for removing single electron pulses (including long electron trains caused by late-extracted electrons from large S2s, or unassociated with energy depositions~\cite{ZIIse,PeterXe10S2only,ZIIIse}); and the fiducial cuts. The fiducial volume, utilizing the effect of self-shielding, is defined so as to exclude the high-background regime at the edges. The dominant background is the PMT radioactivity~\cite{LUXBGPaper2013Malling}.

\begin{figure}[h]
\begin{center}
\includegraphics[width=0.99\textwidth,clip]{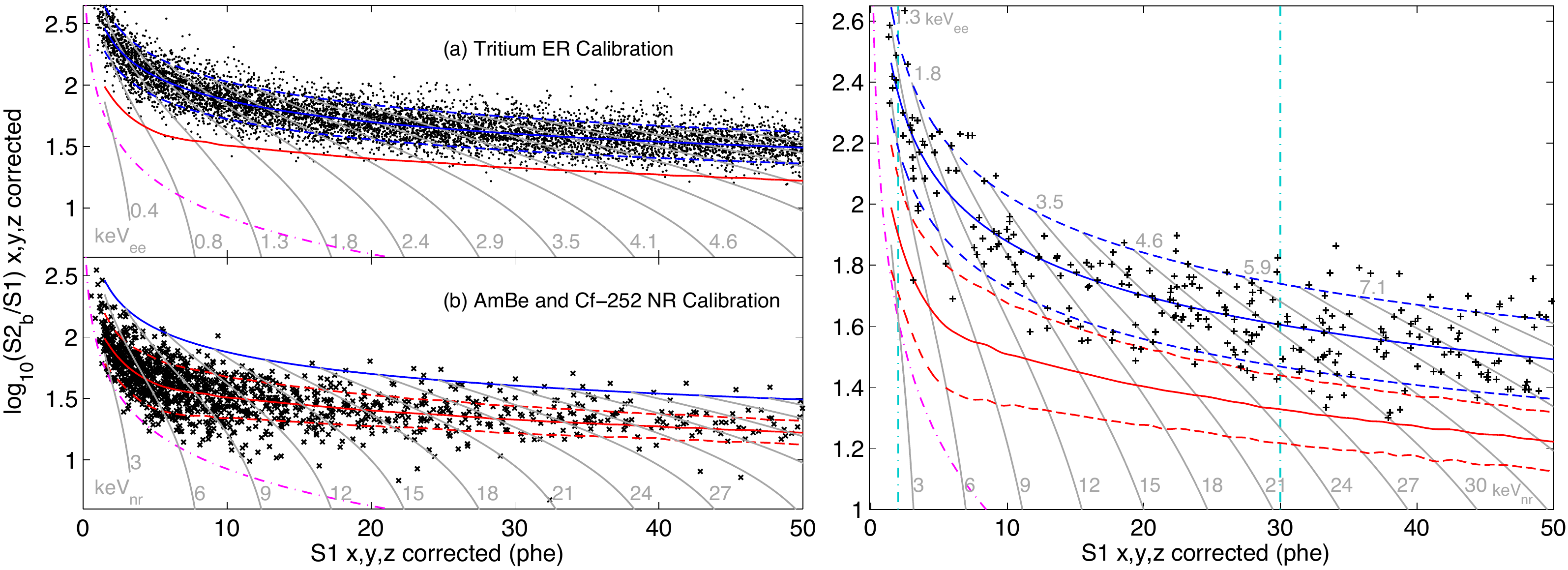}
\caption[]{\label{ERNRbands} {\it Left}: Calibrations of detector response in the 118~kg fiducial volume. The ER (tritium, panel $a$) and NR (AmBe and $^{252}$Cf, panel $b$) calibrations are depicted, with the means (solid line) and $\pm 1.28 \sigma$ contours (dashed lines). This choice of band width (indicating 10\% band tails) is for presentation only. Panel $a$ shows fits to the high statistics tritium data, with fits to simulated NR data shown in panel $b$, representing the parameterizations taken forward to the profile likelihood analysis. The ER plot also shows the NR band mean and vice versa. Gray contours indicate constant energies using an S1--S2 combined energy scale (same contours on each plot). The dot-dashed magenta line delineates the approximate location of the minimum S2 cut. {\it Right}: LUX WIMP signal region.  Events in the 118~kg fiducial volume during the 85.3~live-day exposure are shown.  Lines are as shown on the left, with vertical dashed cyan lines showing the 2--30~phe range used for the signal estimation analysis.}
\end{center}
\end{figure}

The low-energy ER calibration for the region of interest for the background data was performed using a novel tritium source, which provided a high-statistics, homogeneous distribution of energy depositions from betas within the TPC (Figure~\ref{ERNRbands}, top left). Tritiated methane was dispersed $in~situ$ and subsequently removed by purification, making it possible to return to low-background running mode. For NR, AmBe and $^{252}$Cf were used for calibration (Figure~\ref{ERNRbands}, bottom left). In addition to low-energy WIMP-like interactions, these sources are known to create ``neutron-X" events, where a neutron multiply scatters, once in the active region and at least once in an S2-insensitive region, such as below the cathode~\cite{ZIII,Xe1002011}. Such events result in one S1, but also only one S2. (The S1 timescale is too short to allow for two distinguishable pulses in time from a double scatter, though we have used the hit pattern of S1 to identify/reject neutron-X events.) The over-large S1 from the multiple energy depositions causes a significantly lower-than-average log$_{10}$(S2/S1). A further asymmetry, in the other direction, is caused by an ER component. Both AmBe and $^{252}$Cf sources produce gammas in addition to neutrons. Simulating these complexities (inapplicable to WIMP scattering), agreement between the neutron data and the full neutron simulations is achieved. Therefore, after the simulation program was vetted against neutron data, separate simulations were run for WIMPs for the limit calculation, for individual WIMP masses. These were simulations of single scatter NR with appropriate recoil spectra.

The mean, raw leakage of ER events below the mean in the band of NR events in log$_{10}$(S2/S1) versus S1 space (defined by a simulated NR spectrum flat in energy to represent a generic band) was 0.4$\pm$0.1\% in the 2--30~phe S1$_{c}$ range ($c$ meaning position-corrected, as defined earlier); this is a discrimination of 99.6\%. This was defined by looking at all ER events falling below a power law fit to a series of Gaussian means for the NR band in S1$_{c}$ slices. However, since a PLR (Profile Likelihood Ratio) approach~\cite{Cowan} was used for the limit (which utilized a parameterization of the ER band directly from the tritium data without a need for first-principles simulation as for the NR case), instead of the more traditional cut-and-count method, which uses a fixed, linear border in log$_{10}$(S2/S1) to define a WIMP search regime, this leakage value was not used directly in the calculation. We quote it here to illustrate the goodness of separation between the ER and NR bands using a traditional metric. This result, of such low leakage, indicates that light collection may be as important as field for discrimination~\cite{ZIII,DahlThesis}. Furthermore, a comparison of the raw leakage in terms of number of events to that derived from Gaussian fits to the ER band indicates that the ER band is close to Gaussian at a statistical level not previously tested~\cite{Xe1002011}.

The S1$_{c}$ analysis range of 2--30~phe is approximately 3--25~\keVnr, approximate only because of energy channeled into S2, and energy resolution. The upper end was chosen early on, in order to avoid the $\sim$5~keV of energy deposited by cosmogenic $^{127}$Xe activity. In spite of the low NR scintillation yield assumed, the analysis threshold is the lowest ever for a xenon TPC~\cite{ZIII,Xe100,Xe1002011}. At 2~phe LUX still has $\sim$80\% S1 pulse finding efficiency and overall efficiency, a figure cross-checked with different data sets and methods as described earlier. For NR, the conversion to ``electron equivalent" energy or \keVee~is defined in a field-independent fashion as the energy for which ER would produce the same total number of quanta (S1 and S2) as the NR in question. We estimate 3--25~\keVnr~to be 0.9--5.3~\keVee. In the past this conversion was performed using S1 alone, based on the scintillation of $^{57}$Co~\cite{ZIII,Xe100,Xe1002011,Plante2011,Manzur2010}. That failed to account for deposited energy expressed as S2, with the change in ratio of S2 to S1 with field causing the energy definition to be electric field dependent. In addition, the older definition did not account for the energy dependence of ER yield, assuming $^{57}$Co was representative of all ER, long known not to be the case~\cite{Obo}, as recently confirmed again~\cite{Baudis2013}. Using a combined energy scale avoids these issues by looking at the field- and energy-independent total yield summed over S1 and S2. Field and energy change the relative distribution of quanta between the two populations. Unlike for NR, where it is a conversion, for ER \keVee~is meant to represent the best reconstruction of the original energy of a recoiling electron stemming from a gamma or beta.

\section{Result: Dark Matter WIMP Limit}

\begin{figure}[h]
\begin{center}
\includegraphics[width=0.97\textwidth,clip]{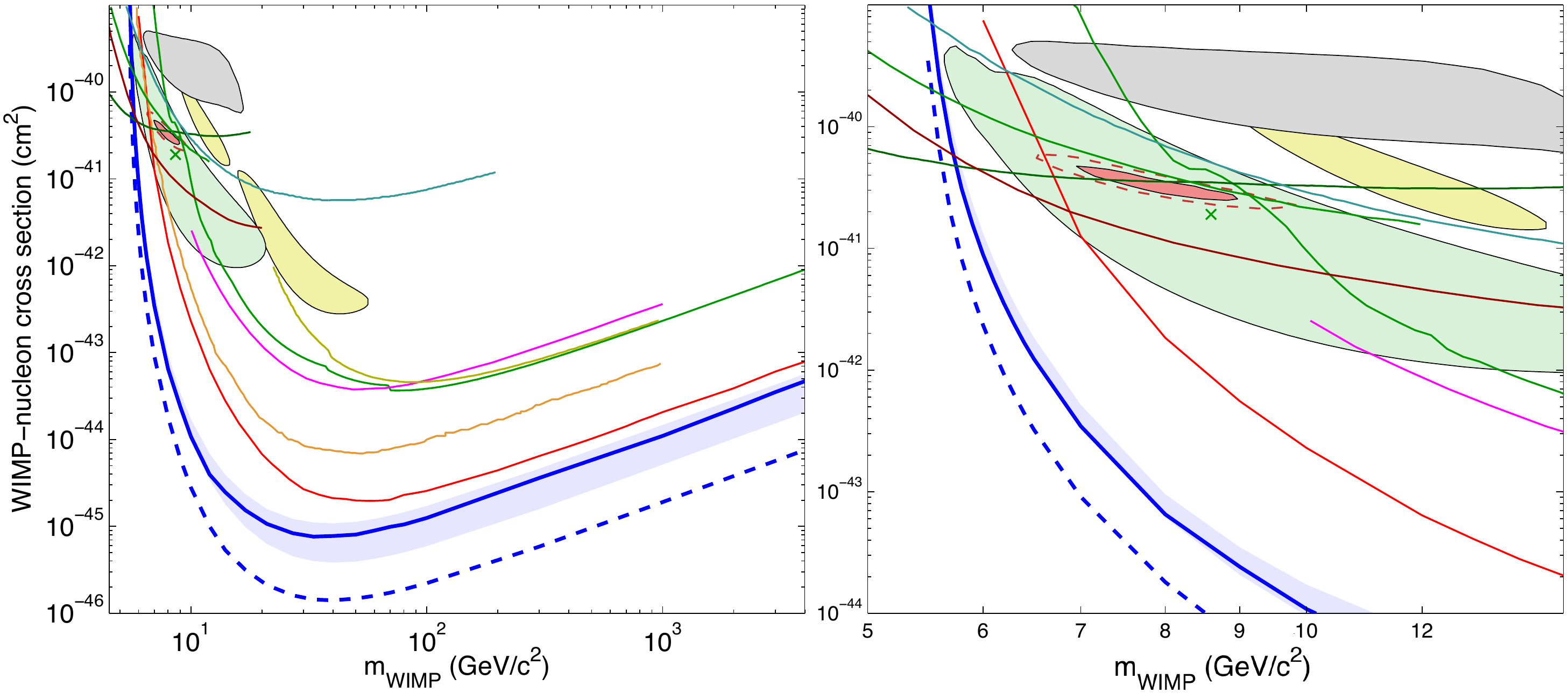}
\caption{\label{limitPlot} {\it Left}: The LUX 90\% confidence limit on the spin-independent elastic WIMP-nucleon cross-section (solid blue line) and 300-day projected sensitivity (dashed blue), together with the ${\pm}1\sigma$ variation from repeated trials, where trials fluctuating below the Poisson limit from a background-free experiment recording zero counts are forced to 2.3 events (blue shaded). We also show Edelweiss~II~\cite{edelweiss} (dark yellow line), CDMS~II~\cite{CDMSLatestResults} (green line), ZEPLIN-III~\cite{ZIII} (magenta line), CDMSlite~\cite{CDMSlite} (dark green line), XENON10 S2-only~\cite{PeterXe10S2only} (brown line), SIMPLE~\cite{SIMPLE} (light blue line), and XENON100 100~live-day~\cite{Xe1002011} (orange line) and 225~live-day~\cite{Xe100} (red line) results. {\it Right} (zoom): In addition to the LUX limit, the regions measured from annual modulation in CoGeNT~\cite{CoGENTLatestResults} (light red, shaded) are shown, along with exclusion limits from low threshold re-analysis of CDMS~II data~\cite{CDMS_2keV} (upper green line), 95\% allowed region from CDMS~II silicon detectors~\cite{CDMS_SiResults} (green shaded) and centroid (green X), 90\% allowed region from CRESST II~\cite{CRESST} (yellow shaded), and DAMA/LIBRA allowed region~\cite{DAMA}, as interpreted by~\cite{DAMA_Savage} (gray shaded).}
\end{center}
\end{figure}

The total number of events was only 160 in 85.3~live-days and 118~kg (Figure~\ref{ERNRbands}, right). The distribution of those events is completely consistent with the ER background, both in log$_{10}$(S2/S1) as well as in terms of position in the volume. 0.64$\pm$0.16 background events total below the S1$_{c}$-dependent NR band centroid were expected for the initial LUX exposure~\cite{LUXBGPaper2013Malling}, and 1 was observed, very close to that border. However, this event is not automatically considered a ``WIMP candidate" because a binary decision is not implemented in terms of log$_{10}$(S2/S1) to determine whether an event is NR or ER, but rather the PLR assigns a continuous probability as a function of S1, S2, radius, and $z$, based on the distributions of the backgrounds in these variables. This method increases statistical power by not discriminating between interaction types with a simplistic binary cut. In keeping with the combined-energy philosophy, after vetting the WIMP Monte Carlo with the empirical NR band, S1 and S2 distributions were simulated together from the theoretical true recoil energy spectrum of each WIMP mass (5.5 to 5000~GeV/$c^{2}$) and input into the PLR calculation, since the energy from NR goes into both S1 and S2. The resulting 90\% C.L. upper limit on the spin-independent WIMP-nucleon interaction cross-section~\cite{PRL} significantly supersedes previous direct detection results~\cite{ZIII,Xe100, CDMSLatestResults, CoGENTLatestResults}, with a minimum occurring at 33~GeV/$c^{2}$ WIMP mass for a cross-section of $7.6 \times 10^{-46}$~cm$^{2}$ (Figure~\ref{limitPlot}).

The LUX limit is much lower, especially at low masses, than the results of past xenon-based experiments, despite the lower NR light yield assumed, because of a higher S1 light collection efficiency, lower S1 threshold (2~phe, as opposed to 3 or 4~phe), and comparable S2 threshold ($\sim$8 extracted electrons, or $\sim$12 original)~\cite{Sz_IDM,ZIII,Xe100,Xe1002011}. For the standard isothermal galactic halo model of Maxwellian-distributed WIMPs and standard astrophysical assumptions, and a generic, non-isospin-violating interacting WIMP particle~\cite{LandS,Savage,McCabe}, LUX is in disagreement with all of the 90--99\% C.L. regions of interest defined by both recent and older experiments that have observed potential low-mass WIMP events~\cite{CoGENTLatestResults,DAMA,CRESST,CDMS_SiResults}. Lower-mass WIMPs not only produce lower-energy recoils than at higher masses but also appear lower in the log$_{10}$(S2/S1)-S1 plane. Detection of such WIMPs would be due exclusively to upward fluctuation in S1 (both intrinsic fluctuation as well as statistical light collection)~\cite{Sorensen2012}, because even with the high light collection in LUX the expectation value for the S1 from a low enough mass WIMP would be below the 2~phe threshold. This makes these WIMPs appear further from the ER band, allowing for good low-mass WIMP sensitivity in xenon even with the falling detection efficiency at low energies of recoil.

A cut-and-count cross-check was performed with the same boundaries in S1$_{c}$ as for the PLR. Excellent agreement was demonstrated. The 0 and 1 background-event cases bound the PLR, which lies closer to the zero background one.~That is what one would expect from a PLR analysis, since it takes the likelihood of an event being NR into account instead of making a rectangular cut. Feldman-Cousins statistics~\cite{FeldCous} was utilized, applying an expectation value of 0.64 background events. The limit with the assumption of 1 background event observed is likely a good upper bound for the PLR because of the 1 event below the NR centroid near 3~phe.

An improvement of a factor of $\sim$5 is expected over the present limit for the upcoming 300-day run, which will be a blind one. The improvement factor is greater than just the ratio of exposures between the second and first runs because of a reduced cosmogenic background. All cosmogenically activated isotopes, including $^{127}$Xe, which produces 1~keV ER unavoidably within the WIMP search region, will have completely decayed away.

\section{Conclusions}

LUX has the largest exposure of any xenon TPC, as well as the lowest threshold. A new internal ER calibration source, tritium, was successfully implemented, and low-energy NR data agree with Monte Carlo, with the location of the NR band well predicted in terms of absolute charge and light yields for an electric field not studied previously. LUX has achieved the most stringent WIMP limit across a wide range of masses. In spite of assumptions more conservative than have been used in the past for xenon detectors, but which we have shown to agree well with the LUX data, our result is in conflict with low-mass WIMP interpretations of signals seen in DAMA, CoGeNT, CRESST, and CDMS Si. With a 300-day run LUX will probe cross-section versus mass parameter space previously unexplored by any other direct detection experiment, with a significantly improved sensitivity compared to this initial result.

\Acknowledgements

This work was partially supported by the U.S. Department of Energy (DOE) under award numbers DE-FG02-08ER41549, DE-FG02-91ER40688, DE-FG02-95ER40917, DE-FG02-91ER40674, DE-NA0000979, DE-FG02-11ER41738, DE-SC0006605, DE-AC02-05CH11231, DE-AC52-07NA27344, DE-FG01-91ER40618; the National Science Foundation under award numbers PHYS-0750671, PHY-0801536, PHY-1004661, PHY-1102470, PHY-1003660, PHY-1312561, PHY-1347449; the Research Corporation grant RA0350; the Center for Ultra-low Background Experiments in the Dakotas (CUBED); the South Dakota School of Mines and Technology (SDSMT). LIP-Coimbra acknowledges funding from Funda\c{c}\~{a}o para a Ci\^{e}ncia e Tecnologia (FCT) through project-grant CERN/FP/123610/2011. Imperial College and Brown University thank the UK Royal Society for travel funds under the International Exchange Scheme (IE120804). The UK groups also acknowledge the institutional support from Imperial College London, University College London, Edinburgh University, and the Science~\&~Technology Facilities Council for Ph.\,D.\, studentship ST/K502042/1 (AB). The University of Edinburgh is a charitable body, registered in Scotland, with registration number SC005336.  This research was conducted using computational resources and services at the Center for Computation and Visualization, Brown University.

We acknowledge the work of the following engineers who played important roles during the design, construction, commissioning, and operation phases: S.~Dardin from Berkeley; B. Holbrook, R. Gerhard, and J. Thomson from University of California, Davis; and G. Mok, J. Bauer, and D. Carr from Lawrence Livermore National Laboratory.
We gratefully acknowledge the logistical and technical support and the access to laboratory infrastructure provided to us by the Sanford Underground Research Facility (SURF) and its personnel at Lead, South Dakota. SURF was developed by the South Dakota Science and Technology Authority, with an important philanthropic donation from T. Denny Sanford, and is operated by the Lawrence Berkeley National Laboratory for the Department of Energy, Office of High Energy Physics.

\end{document}